\documentclass[prb,twocolumn,showpacs,superscriptaddress]{revtex4}

\usepackage{amssymb}

\usepackage{epsfig}

\usepackage{amsmath}

\begin{document}

\title{Exact spontaneous plaquette ground states for high-spin ladder models}

\author{Shu Chen,$^{1}$ Congjun Wu,$^{2}$ Shou-Cheng Zhang,$^{2}$ and Yupeng Wang$^{1}$ \\
${}^{1}$~Institute of Physics, Chinese Academy of Sciences,
Beijing 100080, China\\
${}^{2}$~Department of Physics, Stanford University, Stanford CA
94305-4045}
\date{\today}

\begin{abstract}
We study the exchange physics in high spin Mott insulating systems
with $S=3/2$ which is realizable in ultracold atomic systems. The
high symmetry of $SO(5)$ or $SU(4)$  therein renders stronger
quantum fluctuations than the usual spin-1/2 systems. A
spontaneous plaquette ground state without any site and bond spin
orders is rigorously proved in a ladder spin-3/2 model, whose
topological excitations exhibit fractionalization behavior. The
generalization to the $SU(N)$ plaquette state is also
investigated.
\end{abstract}
 \maketitle

There has been considerable interest recently in high spin systems
with enlarged symmetry among both  condensed-matter and atomic
physics \cite{wu2003,wu2004,Yip,Li98,harada2003,affleck1991}. With
the rapid progress in the field of ultra-cold atomic physics
\cite{Greiner}, optical traps and lattices open up new
possibilities of simulating and manipulating high spin physics
experimentally in a controlled way. For example, the high spin
fermionic $^{40}$K gas has been produced in one-dimensional
optical lattices \cite{Modugno03}. On the other hand,  as a
paradigm in the low-dimensional magnetic systems, the spin chains
and ladders have stimulated intense investigation recently of how
to simulate them using cold atoms. Several schemes have been
proposed to implement these quantum spin models in optical
lattices \cite{Yip,Duan,Cirac04}. So far most of these proposals
are still concentrated in the spin-1/2 and spin-1 systems.
Although most high spin systems only exhibit the usual spin
$SU(2)$ symmetry, a generic high symmetry of $SO(5)$,  or
isomorphic $Sp(4)$, has been rigorously proven in spin 3/2 systems
\cite{wu2003,wu2004}. They may be realized with several candidate
atoms, such as $^{132}$Cs, $^9$Be, $^{135}$Ba, $^{137}$Ba. This
symmetry sets up a framework to unify many seemingly unrelated
properties of Fermi liquid theory, Cooper pairs, and magnetic
structures in such systems. Conventionally, a high spin system is
assumed in the large-$S$ limit, and thus is considered as more
classical-like than its low spin counterpart. However, due to
their high symmetry, the spin-3/2 systems are actually in the
large-$N$ limit. As a result, the quantum fluctuations are
stronger than the usual spin 1/2 systems, which results in many
novel properties.

In this work, a spin-3/2 exchange model with intrinsic $SO(5)$
symmetry is proposed by us for the first time, and it includes the
$SU(4)$ model as its special case. We then study a class of
solvable spin-3/2 ladder models which exhibit the exact
spontaneous plaquette ground states without any site and bond spin
orders, and the topological excitations are studied. To the best
of our knowledge, the existence of a plaquette phase has never
been exactly proved before, therefore our results also provide a
firm ground for understanding the plaquette phase
\cite{Li98,Ueda,Mila01}. The generalization of our theory to
arbitrary spin systems with $SU(N)$ symmetry is also discussed.

We first derive the general $SO(5)$ exchange model from a spin-3/2
Hubbard model in the strong repulsive interaction limit $U_0,
U_2\gg t$, where $U_{0,2}$ are the on-site Hubbard repulsion in
the singlet and quintet channels, respectively. The projection
perturbation theory is employed to study the low energy exchange
process through virtual hopping at quarter-filling, i.e. one
particle per site. For two neighboring sites, the total spin can
be $S_{tot}=0,1,2,3$. Exchange energies in the singlet and quintet
channels are $J_0= 4t^2/U_0,J_2= 4t^2/U_2$ respectively. These two
channels also form $Sp(4)$'s singlet and quintet. On the other
hand, no exchange energies exist in the triplet and septet
channels, which together form a 10D representation of the $Sp(4)$
group. The effective Hamiltonian can be expressed through the bond
projection operators in the singlet channel $Q_0(ij)$ and quintet
channel $Q_2(ij)$ as $H_{ex}=\sum_{\langle i,j\rangle
}h_{ij}=-\sum_{\langle i,j\rangle }\big
\{J_0Q_0(i,j)+J_2Q_2(i,j)\big \},$ or in an explicitly $Sp(4)$
invariant form as
\begin{equation}
H_{ex}=\sum_{\langle i,j\rangle }\big \{ \frac 14\left[
c_1A^{\gamma_1} (i)A^{\gamma_1} (j)+c_2A^{\gamma_2}
(i)A^{\gamma_2} (j)\right] -c_3\big \} \label{ex}
\end{equation}
with $c_1=\frac{J_0+J_2}4,$ $c_2=\frac{3J_2-J_0}4$ and
$c_3=\frac{J_0+5J_2}{16},$ where $A^{\gamma_1} =2L_{ab}=c_\alpha
^{\dagger }\Gamma _{\alpha \beta }^{ab}c_\beta ~(\gamma_1
=1,\cdots ,10,~1\le a<b\le 5)$ and $A^{\gamma_2} =2n_a=c_\alpha
^{\dagger }\Gamma _{\alpha \beta }^ac_\beta ~(\gamma_2 =11,\cdots
,15,~1\le a\le 5)$ with the Dirac $\Gamma $ matrices which could
be found in Ref. \cite{wu2003}. The $n_a$ operators transform as a
5-vector under the $Sp(4)$ group and the $L_{ab}$ operators form
the 10 generators of the $SO(5)$ group.

It is obvious that an $SU(4)$ symmetry appears at $J_0=J_2$. Then
Eq. (\ref{ex}) reduces into the $SU(4)$ symmetric form with the
fundamental representations on every site where $L_{ab}$ and $n_a$
together (or $A^\gamma $ with $\gamma =1,\cdots,15$) form the 15
generators of the $SU(4)$ group. In this case, the spin 3/2
Hamiltonian is equivalent to the well-known $SU(4)$ spin-orbital
model up to a constant term\cite{Li98,Ueda}. Equation (\ref{ex})
can be expressed by the usual $SU(2)$ spin operators with
bi-quadratic and bi-cubic terms as
\begin{equation}
H_{ex}^{\prime }=\sum_{\langle i,j\rangle}a~(\vec{S}_i\cdot
\vec{S}_j)+b~(\vec{S}_i\cdot \vec{S}_j)^2+c~(\vec{S}_i\cdot
\vec{S}_j)^3, \label{ex2}
\end{equation}
with $a=-\frac 1{96}(31J_0+23J_2),~~b=\frac 1{72}(5J_0+17J_2)$ and
$c=\frac 1{18}(J_0+J_2)$.
\begin{figure}[tbp]
\includegraphics[width=3.5in]{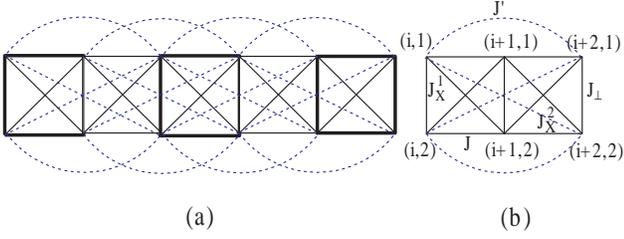}
\caption{(a)The spin ladder model. Plaquettes in bold lines
represent the plaquette singlets. (b)The basic block to construct
the spin ladder model.} \label{fig1}
\end{figure}

The magnetic structure of the one-dimensional spin-3/2 model
(\ref{ex2}) can be analyzed by bosonization methods combined with
renormalization group methods via studying the corresponding
Hubbard model at quarter-filling \cite{affleck1988}. A spin gap
opens with the appearance of the spin Peierls order at
$U_2>U_0>0$, while it varnishes at $U_0 \ge U_2>0 $ \cite{wu2004},
where $U_{0,2}$ are the on-site repulsions in the spin singlet and
quintet channels respectively.

Now we construct a spin-3/2 ladder model exhibiting an exact
plaquette ground state, which is an $SU(4)$ generalization of the
valence bond state (VBS) as shown in Fig 1. Different from the
usual $SU(2)$ case where every bond is an $SU(2)$ singlet, here at
least four sites, i.e. a plaquette, are needed to form an $SU(4)$
singlet, thus there is only plaquette order without any bond order
in this state. As shown in Fig. 1(a), the model is described by
the Hamiltonian
\begin{eqnarray}
H_{lad} &=&J_{\perp }\sum_ih_{(i,1),(i,2)}+J\sum_{i,\alpha
}h_{(i,\alpha
),(i+1,\alpha )}  \nonumber \\
&&+\sum_i\sum_{\delta =1}^2J_{\times }^\delta \left[
h_{(i,1),(i+\delta
,2)}+h_{(i,2),(i+\delta ,1)}\right]  \nonumber \\
&&+J^{\prime }\sum_{i,\alpha }h_{(i,\alpha ),(i+2,\alpha )},
\label{SO5ladder}
\end{eqnarray}
where the site is labelled by its rung number $i$ and chain index
$\alpha =1,2$ and $h_{ij}=\frac 14\left[ c_1A^{\gamma
_1}(i)A^{\gamma _1}(j)+c_2A^{\gamma _2}(i)A^{\gamma _2}(j)\right]
$. Here we take $J_{\perp },$ $J,$ $J_{\times }^1,$ $J_{\times
}^2$ and $J^{\prime }$ as positive constants in unit of $J_0$ and
set $J=1$. The regular railway ladder model corresponds to
$J_{\times }^1=J_{\times }^2=J^{\prime }=0$. To gain some
intuition about the plaquette state, we first consider a 4-site
system with diagonal exchanges (a tetrahedron). For such a simple
system with $SU(4)$ symmetry, the GS is a plaquette
singlet\cite{Li98} defined as
\[
su_4\left( 1234\right) =\sum_{\mu \upsilon \gamma \delta }\frac
1{\sqrt{24} }\Gamma _{\mu \upsilon \gamma \delta }\left| 1_\mu
2_\nu 3_\gamma 4_\delta \right\rangle
\]
where $\Gamma _{\mu \upsilon \gamma \delta }$ is an antisymmetric
tensor and $\mu ,\upsilon ,\gamma ,\delta =\pm \frac 32,\pm \frac
12.$ Such an $SU(4)$ singlet is rotationally invariant under any
of the fifteen generators $A^i$ of an $SU(4)$ group with
$i=1,\cdots ,15.$ For a tetrahedron with $SO(5)$ symmetry, the
$SU(4)$ singlet is no doubt an eigenstate with the eigen-energy
$e_{\boxtimes }=-5c_1-\frac 52c_2$. Furthermore, it is easily
verified that the $SU(4)$ singlet is the ground state of the
four-site $ SO(5) $ exchange model for $c_2\ge 0$. \\

We now focus on the SU(4) ladder described by Eq.
(\ref{SO5ladder}) with $ c_2=c_1$. It turns out that the exact
ground state of the SU(4) ladder model is a doubly degenerate
singlet provided the relation
\begin{equation}
J_{\perp }=\frac 32J=\frac 32J_{\times }^1=3J_{\times
}^2=3J^{\prime } \label{constraint1}
\end{equation}
is fulfilled. The degenerate GSs are composed of products of
nearest-neighboring plaquette singlets. Explicitly, for a ladder
with length $M$, the two degenerate ground states are given by
\begin{eqnarray*}
\left| P_1\right\rangle &=&\prod_{i=1}^{M/2}su_4\left[
(2i,1),(2i,2),(2i+1,1),(2i+1,2)\right] \\
\left| P_2\right\rangle &=&\prod_{i=1}^{M/2}su_4\left[
(2i-1,1),(2i-1,2),(2i,1),(2i,2)\right]
\end{eqnarray*}
with the corresponding ground energy
\begin{equation}
E=-M\frac{35}8c_1,  \label{GSE}
\end{equation}
where periodic boundary condition is assumed $M+1\equiv 1.$

The proof of the above conclusion can be understood through two
steps. First, one observes that the plaquette singlet product is
no doubt an eigenstate of the global Hamiltonian because any
generator of the $SU(4)$ aside from the plaquette acting on the
plaquette singlet is zero, i.e.
\[
A_5^\gamma \left[ A_1^\gamma +A_2^\gamma +A_3^\gamma +A_4^\gamma
\right] su_4(1234)\left| 5\nu \right\rangle =0
\]
where $\gamma =1,\cdots ,15.$ Secondly, we prove that such an
eigenstate is the ground state of the global Hamiltonian. In order
to prove it, we utilize the Rayleigh-Ritz variational principle
\begin{equation}
\left\langle \Psi \right| \sum_{i=1}^Mh(B_i)\left| \Psi
\right\rangle \geq E_{g.s.}\ge \sum_{i=1}^Me_{g.s.}(B_i),
\end{equation}
which implies that an eigenstate $\left| \Psi \right\rangle $ is
the ground state of a global Hamiltonian if it is simultaneously
the ground state of each local sub-Hamiltonian \cite{SS}. Here
$E_{g.s.}$ is the ground state energy of the global Hamiltonian
which is represented as a sum of $M$ sub-Hamiltonians, say,
$H=\sum_{i=1}^Mh(B_i),$ and $e_{g.s.}(B_i)$ represents the ground
state energy of a sub-Hamiltonian $h(B_i)$.

Now we apply the above general principle to study our model given
by Eq.(\ref{SO5ladder}). Explicitly, as long as
Eq.(\ref{constraint1}) holds true, we can decompose
Eq.(\ref{SO5ladder}) as
\begin{equation}
H=\sum_{i=1}^Mh(B_i)  \label{H1},
\end{equation}
where $h(B_i)=J^{\prime }\sum_{\left\langle ij\right\rangle
}h_{ij}$ denotes the Hamiltonian of a six-site block as shown in
Fig.1(b) and $\left\langle ij\right\rangle $ represents all the
available bonds in the block of $ B_i.$ For convenience, we use
the rung index on the left side of a six-site block to label the
block. For such a six-site cluster with 15 equivalent bonds, the
local Hamiltonian can be represented as a Casimir operator and the
representation with smallest Casimir corresponds to the Young
diagram $ \left[ 2^21^2\right] $. It follows that eigenstates
given by
\begin{eqnarray*}
\left| B_i\right\rangle _1 &=&su_4\left[
(i,1),(i,2),(i+1,1),(i+1,2)\right]
\otimes d_{i+2} \\
\left| B_i\right\rangle _2 &=&d_i\otimes su_4\left[
(i+1,1),(i+1,2),(i+2,1),(i+2,2)\right]
\end{eqnarray*}
are two of the degenerate ground states of the six-site block
$B_i$ with the ground energy $e_g(B_i)=-J^{\prime
}(\frac{15}2c_1+\frac 54c_1),$ where $d_i$ represents a dimer on
the $i$th rung
\[
d_i=\left[ (i,1),(i,2)\right] =\frac 1{\sqrt{2}}\Gamma _{\mu \nu
}\left| (i,1)_\mu (i,2)_\nu \right\rangle,
\]
with $\Gamma _{\mu \nu }$ denoting an antisymmetric tensor. A
global eigenstate can be constructed by a combination of local
eigenstates $\left| B_i\right\rangle _1$ or $\left|
B_i\right\rangle _2$ because the dimers on the side of the block
are free in the sense that they can form plaquette singlets with
other sites belonging to the neighboring blocks. Now it is obvious
that the eigenstates $\left| P_1\right\rangle $ and $\left|
P_2\right\rangle $ are the ground state of each sub-Hamiltonian
$h(B_i)$, and therefore the degenerate GSs of the global
Hamiltonian with the GS energy given by Eq.(\ref{GSE}). In fact,
the constraint relation Eq.(\ref{constraint1}) can be further
released to
\begin{equation}
J_{\perp }\geq \frac 32J=\frac 32J_{\times }^1=3J_{\times
}^2=3J^{\prime }, \label{constraint2}
\end{equation}
which means that the degenerate plaquette products are the GSs
even in the strong rung limit $J_{\perp }$ $\gg J.$ Furthermore,
we note that $\left| P_1\right\rangle $ and $\left|
P_2\right\rangle $ are the eigenstate of the $ SU(4)$ ladder model
even for arbitrary $J_{\perp }$, but not necessarily the GS.

This plaquette-product state is a spin gapped state with only
short-range spin correlations. The two GSs are spontaneously
tetramerized and thus break translational symmetry. The elementary
excitation above the GS is produced by breaking a plaquette
singlet with a finite energy cost, thus leading to an energy gap.
Two kinds of excitations are possible, either a magnon-like
excitation or a spinon-like excitation. The gap size of a
magnon-like excitation is approximately equivalent to the energy
spacing level between the GS and the first excited state of a
local tetrahedron. For the model (\ref{constraint2}), the first
two excited states above the plaquette singlet are represented by
the Young diagrams $\left[ 2^11^2\right] $ and $\left[ 2^2\right]
$ and the corresponding gap sizes are $\Delta _m=4c_1$ and $\Delta
_s=6c_1$ respectively. The spinon-like excitation is made up of a
pair of rung dimers which propagate along the leg to further lower
the energy, and thus the excitation spectrum is a two-particle
continuum. The propagating dimer pairs behave like domain-wall
solitons (kink and anti-kink) connecting two spontaneously
tetramerized ground state. We represent an excited state with a
kink at site $2m-1$ and an anti-kink at site $2n$ as $\Psi \left(
m,n\right) $ , so the corresponding momentum-space wavefunction is
\[
\Psi \left( k_1,k_2\right) =\sum_{1\leq m\leq n\leq
M}e^{i(2m-1)k_1+i2nk_2}\Psi \left( m,n\right) .
\]
The excited energy can be calculated directly by using the above
variational wavefuction\cite{SS}. For a spontaneously tetramerized
ladder system, it is reasonable to assume that the kink and
antikink are well separated because there exists no intrinsic
mechanism of binding them together to form a bound state,
therefore we treat kink and antikink separately. Since the state $
\Psi \left( m\right) $ is not orthogonal with the inner product
given by $ \left\langle \Psi \left( m^{\prime }\right) \right|
\left| \Psi \left( m\right) \right\rangle =\left( \frac 16\right)
^{\left| m^{\prime }-m\right| },$ thus $\Psi \left( k_1,k_2\right)
$ has a nontrivial norm. After considerable but straightforward
algebra, we get
\[
\varepsilon \left( k_1,k_2\right) =\frac{37}{35}\Delta -\frac
6{35}\left( \cos 2k_1+\cos 2k_2\right) \Delta ,
\]
where $\Delta =5c_1$ is the energy level spacing between the GS
and the excited state where a local plaquette is broken into two
rung singlets. Since the SU(4) spin-3/2 model can be mapped into
the spin-orbital model, our above results can be directly applied
to the corresponding spin-orbital ladder model. For an SU(4)
spin-orbital railway-ladder model, we noticed that it has similar
degenerate GSs composed of $SU(4)$ singlet from the work of
Bossche {\it et al.} based on the exact diagonalization method and
analytical analysis in strong coupling limits \cite{Mila01}.

For the SO(5) ladder (\ref{SO5ladder}) with $c_2>0$, one might
expect that the degenerate plaquette states are the GSs in the
parameter regime (\ref {constraint2}) like in the case of the
SU(4) ladder. Unfortunately, one can check that those plaquette
states are no longer the eigenstates and therefore not the exact
GSs. Nevertheless, we can prove that $\left| P_1\right\rangle $
and $\left| P_2\right\rangle $ are degenerate eigenstates of the
global Hamiltonian if $J_{\perp }=J=J_{\times }^1=2J_{\times
}^2=2J^{\prime }.$
\begin{figure}[tbp]
\includegraphics[width=0.6\linewidth]{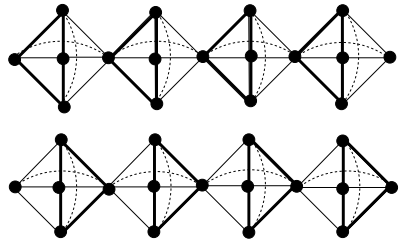}
\caption{Two-fold degenerate ground state of the $SU(4)$
diamond-chain model. The Heisenberg exchange is defined on every
linked bond with the equal magnitude.} \label{fig2}
\end{figure}

Next we turn to a solvable diamond-chain model as shown in Fig.2,
where the spin-3/2 Heisenberg model is defined on the linked bonds
with the same exchange energy $H=J\sum_{\langle ij\rangle}
h_{ij}.$ For an SU(4) model, the Hamiltonian can be written as a
sum of the Casimir of the total spin in each five-site cluster.
Thus we conclude that the GSs are doubly degenerate plaquette
singlets because among the possible representations composed by
5-sites, the above ground state configurations provide the
smallest Casimir. By the same reasoning as above, this conclusion
holds true for the SO(5) case with $c_2 >0$.  The excitation is
composed of a spinon and a three-site bound state as a result of a
plaquette singlet breaking up to a 1+3 pattern, which bears some
similarity to the above spin ladder model where symmetric
spinon-like excitations correspond to a plaquette singlet breaking
up to a 2+2 pattern.
\begin{figure}[tbp]
\includegraphics[width=0.8\linewidth]{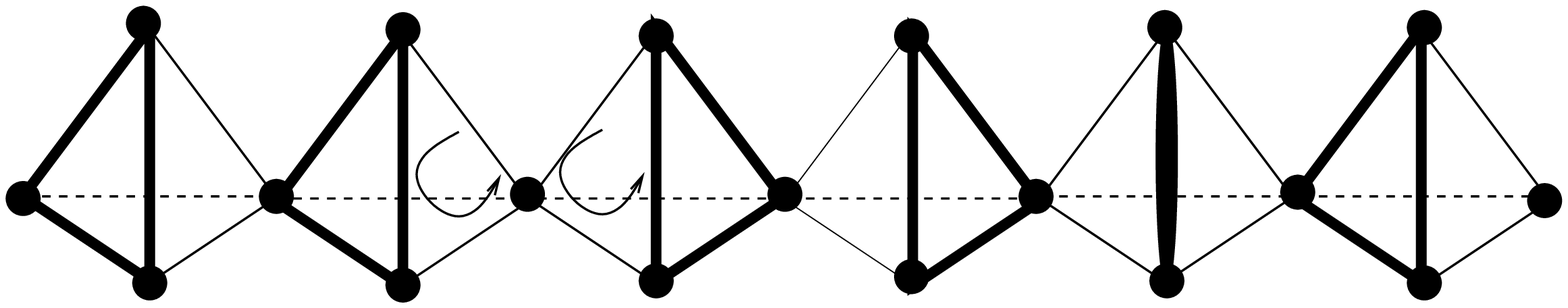}
\caption{ Excitations in the $SU(3)$ model as the nearest-neighbor
Heisenberg model defined on the 1D corner-shared tetrahedrons. }
\label{fig3}
\end{figure}

The exact plaquette GS discussed above can be straightforwardly
generalized to the $SU(N)$ case. The $SU(2)$ case corresponds to
the celebrated spin-1/2 Majumdar-Ghosh
model.\cite{Majumdar70,Chen} The $SU(3)$ case as shown in Fig.
\ref{fig3} is particularly interesting. It is defined in a 1D
correspondence of the pyrochlore lattice where only the nearest
neighbour bond interaction is involved. The Hamiltonian is
\begin{eqnarray}
H=J\sum_{\langle ij\rangle }\sum_{\gamma =1}^8A^\gamma (i)A^\gamma
(j),
\end{eqnarray}
where $A$'s are the eight $SU(3) $ generators in the fundamental
representations, and $\langle ij\rangle $ means the sum over the
nearest neighbors. It can also be written as the spin-1
representation as $H=J\sum_{\langle ij\rangle }\vec{S}(i)\cdot
\vec{S }(j)+a(\vec{S}(i)\cdot \vec{S}(j))^2,$ where the $SU(3)$
point is located at $a=1$. The Hamiltonian can be written as a sum
of the Casimir of the total spin in each tetrahedra. We know that
the representations with the smallest Casimir made out of four
sites in the fundamental representations is three dimensional. The
excitations also have a gap and are fractionalized as in the 1D
polymer systems. The three-site singlet (quark) is broken up to a
1+2 pattern as one monomer in the fundamental representation, and
the other two form an anti-fundamental (anti-quark)
representation. Thus the quark and anti-quark pair states are 3*3
fold degenerate. The singlet site monomer can hop around in the
two faces without breaking more singlets.  The detail will be the
subject of a future publication.

In summary, we derived and studied the effective $SO(5)$
spin-$3/2$ exchange model as well as the spin ladder models with
exact two-fold degenerate plaquette GS. Our results indicate the
formation of spontaneously tetramerized GS for a translational
invariant spin ladder system. Quantitative results for the
elementary excitation spectrum of the $SU(4)$ spin-orbital ladder
are also obtained. Due to the existence of an intrinsic $SO(5)$
symmetry, we expect that the plaquette phase of the spin-3/2
models can be observed in optical lattices in the future
experiment. The generalization of our theory to the system with
$SU(3)$ or even $SU(N)$ symmetry is also addressed.

C. W. thanks D. Arovas for useful discussions. S. C. thanks the
``Hundred Talents" program of CAS and the NSF of China under Grant
No. 10574150 for support. This work is also supported by the NSF
under grant numbers DMR-0342832 and the US Department of Energy,
Office of Basic Energy Sciences under contract DE-AC03-76SF00515.


\begin{thebibliography}{99}
%\begin{references}
\bibitem{wu2003}  C. Wu \textit{et al}, Phys. Rev. Lett. \textbf{91}, 186402
(2003).

\bibitem{wu2004}  C. Wu, cond-mat/0409247.

\bibitem{Yip}  S. K. Yip, Phys. Rev. Lett. \textbf{90},
250402 (2003).

\bibitem{Li98}  Y. Q. Li, M. Ma, D. N. Shi, and F. C. Zhang, Phys. Rev.
Lett. \textbf{81}, 3527(1998)

\bibitem{harada2003}  K. Harada \textit{et al}, Phys. Rev. Lett. \textbf{
90 }, 117203 (2003).

\bibitem{affleck1991}
I. Affleck, D.P. Arovas, J.B. Marston and D. Rabson. Nucl. Phys. B
366, 467 (1991).

\bibitem{Greiner} M. Greiner \textit{et al}, Nature (London) \textbf{
415}, 39 (2003); C. Orzel \textit{et al}, Science \textbf{ 291},
2386 (2003); D. Jaksch \textit{et al}, Phys. Rev. Lett.
\textbf{81}, 3108 (1998).

\bibitem{Modugno03}  Modugno \textit{et al}, Phys. Rev. A ~\textbf{68},
011601 (2003).

\bibitem{Duan} L.-M. Duan, E. Demler and M. D. Lukin, Phys. Rev. Lett. \textbf{91}, 090402
(2003).

\bibitem{Cirac04} J. J. Garcia-Ripoll \textit{et al}, Phys. Rev. Lett. \textbf{93},
250405 (2004).

\bibitem{Ueda} Y. Yamashita, N. Shibata, and K. Ueda, Phys. Rev. B. \textbf{58},
9114 (1998); S. K. Pati, R. R. P. Singh, and D. I. Khomskii, Phys.
Rev. Lett. \textbf{81}, 5406 (1998).

\bibitem{Mila01}  M. vandenBossche, P. Azaria, P. Lecheminant, and F. Mila,
Phys. Rev. Lett. \textbf{86}, 4124 (2001).


\bibitem{affleck1988}  I. Affleck, Nucl. Phys. B \textbf{305}, 582 (1988).

\bibitem{SS}  B. S. Shastry and B. Sutherland, Phys. Rev. Lett. \textbf{47},
964 (1981).

\bibitem{Majumdar70}  C. K. Majumdar and D. K. Ghosh, J. Math. Phys.\textbf{
\ 10}, 1388(1969);D. Sen \textit{et al}, Phys. Rev. B. {\bf 53},
6401 (1996); T. Nakamura and K. Kubo, Phys. Rev. B. {\bf 53}, 6393
(1996).

\bibitem{Chen} S. Chen, H. Buttner and J. Voit, Phys. Rev. B, \textbf{67},
054412, (2003); Phys. Rev. Lett. \textbf{87}, 087205, (2001).



%\end{references}



\end{thebibliography}
\end{document}